# The effects of cosmic rays and solar flares on the IRAC detectors: the first two years of in-flight operation

Joseph L. Hora[*a], Brian M. Patten[a], Giovanni G. Fazio[a], William J. Glaccum[b]

[a]Harvard-Smithsonian Center for Astrophysics, 60 Garden St., MS-65, Cambridge, MA USA 02138;
[b]Spitzer Science Center, MS 220-6, Caltech, Pasadena, CA, USA 91125

## ABSTRACT

The Infrared Array Camera (IRAC) is a four-channel camera on the *Spitzer Space Telescope*, one of three focal plane science instruments. IRAC uses two pairs of 256×256 pixel InSb and Si:As IBC detectors to provide simultaneous imaging at 3.6, 4.5, 5.8, and 8 μm. IRAC experiences a flux of cosmic rays and solar protons that produce transient effects in science images from each of the arrays, with 4 – 6 pixels per second being affected during each integration. During extreme solar flares, IRAC experiences a much higher rate of transients which affects the science data quality. We present cosmic ray rates and observed detector characteristics for IRAC during the first two years of science operation, and rates observed in a period of elevated solar proton flux during an intense solar flare in January 2005. We show the changes to the IRAC detectors observed since launch, and assess their impacts to the science data quality.

**Keywords:** *Spitzer Space Telescope*, IRAC, infrared detectors, cosmic rays, image processing and artifacts, space instrumentation

## 1. INTRODUCTION

The Infrared Array Camera[1,2] (IRAC) is one of three focal plane instruments on the *Spitzer Space Telescope*[3,4]. IRAC is a four-channel camera that obtains simultaneous images at 3.6, 4.5, 5.8, and 8 μm. Two adjacent 5.2′×5.2′ fields of view in the *Spitzer* focal plane are viewed by the four channels in pairs (3.6 and 5.8 μm; 4.5 and 8 μm). All four detector arrays in the camera are 256×256 pixels in size, with the 3.6 and 4.5 μm channels using InSb arrays and the 4.5 and 8 μm channels using Si:As IBC arrays[5,6,7,8,9]. Both detector types have pixels with a size of 30×30 μm, but the Si:As detector material is thicker (50 μm) than in the InSb devices (7 μm).

Launched on August 25, 2003, *Spitzer* was placed into an Earth-trailing solar orbit, which positions the spacecraft well outside any shielding effects of the Earth's magnetosphere or enhancements in charged particle flux due to the Earth's radiation belts[9,11]. Following launch, *Spitzer* underwent a 3-month evaluation during the In-Orbit Checkout (IOC) and Science Verification periods[12], during which the telescope was focused[13,14] and a functional checkout of the instruments and observational templates was performed[2]. Since launch, IRAC has been operated for more than 7000 hours, collecting a wealth of calibration and science data. The camera is now being used regularly as a part of the nominal operations (nomops) phase of the mission.

On orbit, the arrays experience an environment that includes high energy ions which produce transient events in IRAC images. Transient events are where a signal is produced in one or more pixels over a time much shorter than the total exposure time for the image, and are not caused by photons from astronomical sources. The signals are random both spatially and temporally. In general, the transient is not evident in the following exposure; however, for some strong transients, a residual image is present in subsequent frames that fades in a manner similar to residual images from bright astronomical sources. Here we will refer to all transients as "cosmic rays" (CRs), regardless of their origin. In a previous paper[15], we presented the characteristics of CRs as they appear in the IRAC science images. In this paper, we present CR rates and changes in detector properties observed in the period December 2003 through April 2006, as well as during a period of elevated solar proton flux following a series of X-class solar flares in January 2005.

---

[*] jhora@cfa.harvard.edu; phone 1 617 496 7548; fax 1 617 495 7490; cfa-www.harvard.edu/irac

# 2. IRAC ARRAY PERFORMANCE MONITORING

Calibration data are obtained during each IRAC science campaign to monitor the instrument's performance. The data set consists of observations of astronomical standard stars, regions of low and high zodiacal background levels, and frames obtained with the internal flood calibration lamps. The calibration strategy and results have been described elsewhere[16]. Since launch, there has been no detected change to the average instrument responsivity as determined from measurements of astronomical standards. In this paper we will focus on changes to individual pixels since launch that are presumably a result of damage due to radiation or other effects. The detector characteristics tracked are the number of hot (high dark current), dead, or noisy pixels in each array, and the median number of CR-affected pixels per second. Hot pixels are defined as those having dark current > 1 $e^-$/s for the InSb arrays (>10 $e^-$/s for the Si:As arrays), dead pixels are those with a responsivity <50% of the median of nearby pixels, and noisy pixels are those with a standard deviation of twice that of the array median in successive frames of the low background region.

## 2.1. Performance monitoring data set and reduction

The data used for assessment of the number of hot and noisy pixels are a set of 100 sec dark frames (50 sec for channel 4) taken at least twice per campaign. In most cases, only the last set of dark frames in a campaign were used for this analysis. These are close in time to the flood lamp frames that are used to determine the number of "dead" or low responsivity pixels. Measurements from IOC through nom ops campaign 30 (September 2003 – April 2006) were analyzed. For the statistics below, only a 250×250 pixel region is considered (region [3:252,3:252]) to avoid noisy columns and rows along the array edges.

The dark frame data are taken in the "Best NEP dark" region, centered near +17h40m, +69d00m (J2000). This is within the continuous viewing zone (CVZ) and was selected to have no bright sources. The 100 sec AOR performs a 3×3 mapping pattern with 3 large dithers per position. For this analysis, the raw data are used, along with the Basic Calibrated Data (BCD) headers for accurate pointing information. First, a dark frame is made from the median of the data using the imcombine command in iraf. This removes the non-uniform sources and leaves only a background offset level from the sky and the detector dark pattern, which includes dark current and readout pattern. In order to remove the sky emission and large features such as the faint glow in the lower right corner, a new image is constructed with the iraf median command using a 5×5 box size, which leaves only the large-scale features of the image. This median image is subtracted from the background to create a hot pixel image.

The number of noisy pixels is determined from the "sigma" maps created by iraf/imcombine, which is the standard deviation of that pixel calculated when the background frame described above was made. Since there are stars in the field, a source falling on a pixel that is not rejected by the imcombine sigma-clipping algorithm might slightly increase the noise estimate for that pixel. However, since the dark frames are taken at different orientations during the year, a pixel samples a different set of sky positions each campaign. Therefore this effect should not lead to a systematically higher noise for one particular pixel, but might add to the scatter of the noise values by occasionally increasing the standard deviation of a pixel for a campaign. The number of noisy pixels changes from campaign to campaign, and increases are correlated with high proton flux rates during solar flares. However, there is not a large systematic increase in the noisy pixel numbers over the period examined here. The number of CRs are also determined from the 100 sec dark field measurements. A mosaic is constructed of the frames, using the multiple frame coverage to reject transients. Then the individual images are registered and compared to the mosaic, and the number of pixels above a cutoff level (after masking out stars and hot pixels) are counted and divided by the exposure time to yield the number of CR-affected pixels per second. More pixels are affected by each CR in channels 3 and 4, probably due to the thicker detectors, which leads to a slightly higher rate of CR pixels/sec in those channels.

Dead pixels are monitored using the calibration frames obtained with the flood lamps. A low-background field is imaged with the flood lamps off and then on, without dithering. The lamp-off frames are subtracted from the lamp-on frames to remove stars and give the response to the lamp only. Each pixel is compared to the local median in a 5×5 pixel region around it, and those with a value <50% of the median are flagged as dead.

The detector monitoring results are shown in Figure 1– Figure 4. The CR statistics are shown in Figure 5. An example channel 4 hot pixel map is shown in Figure 6. The median values of each of the parameters are summarized in Table 1.

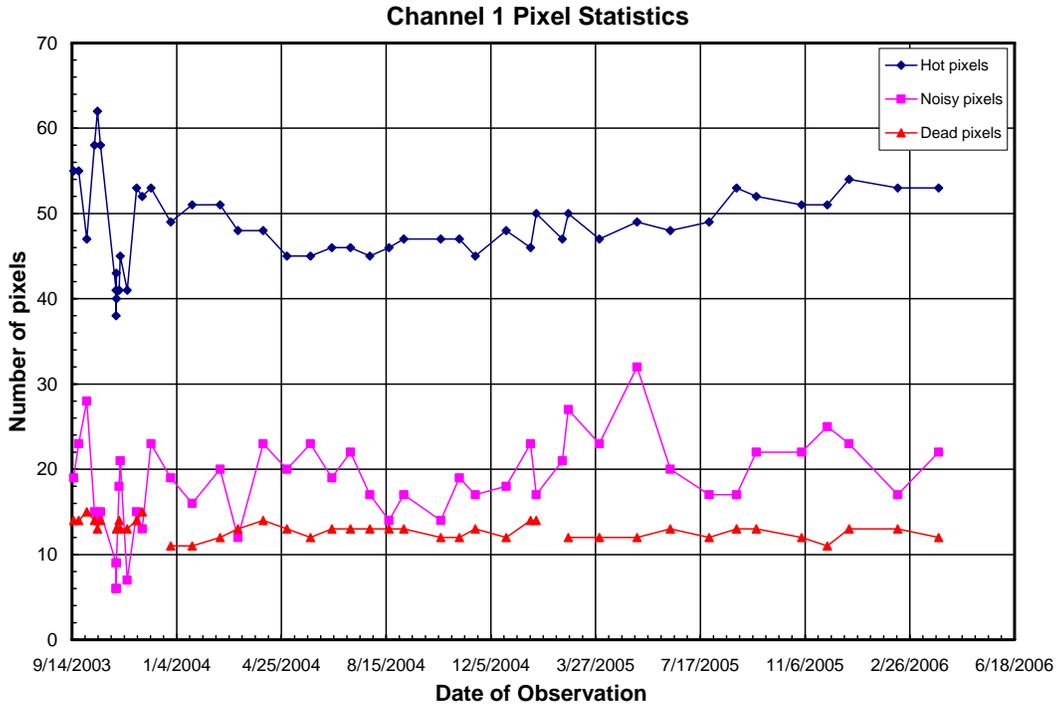

Figure 1. Channel 1 pixel statistics for the mission. The number of hot, dead, and noisy pixels are roughly constant throughout the mission.

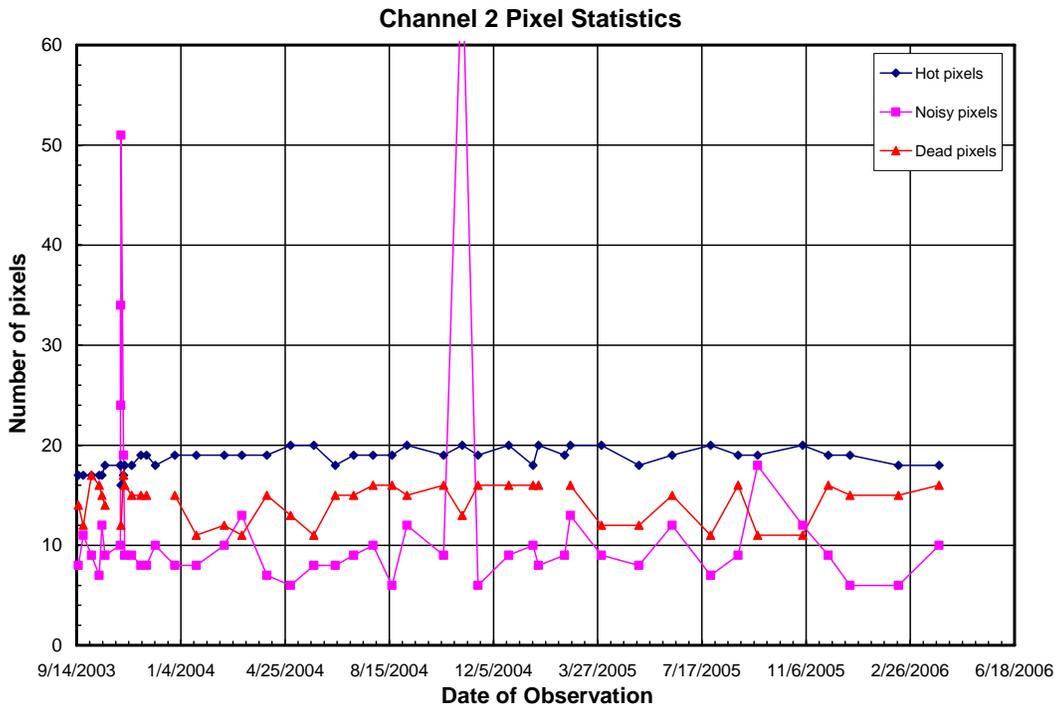

Figure 2. Channel 2 pixel statistics for the mission. Temporary increases in the number of noisy pixels seems to be the only effect of the solar flares.

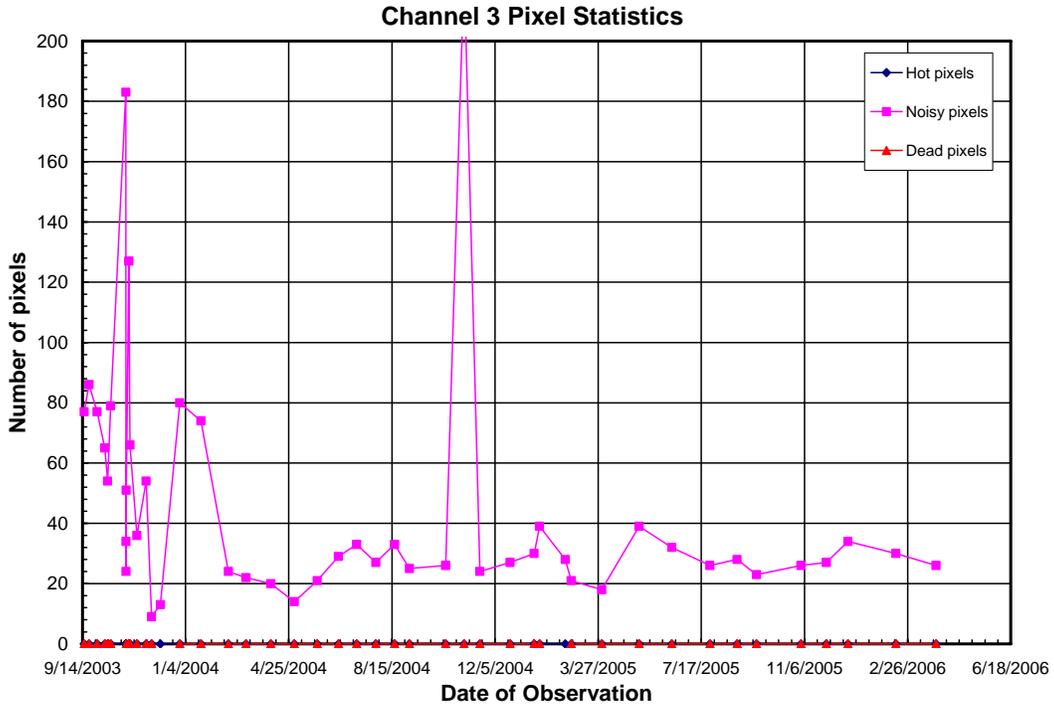

Figure 3. Channel 3 pixel statistics for the mission. There are no hot or dead pixels, and a temporary increase in the number of noisy pixels seems to be the only effect of the solar flares.

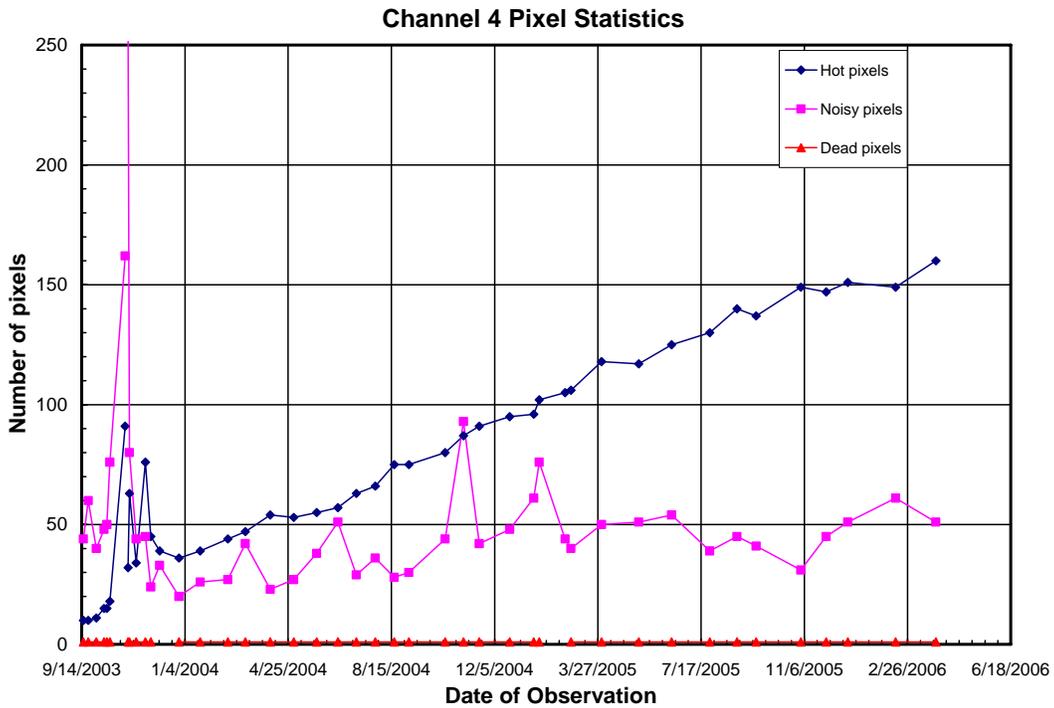

Figure 4. Channel 4 pixel statistics for the mission. There is one dead pixel. The effects of the flares seem to be a temporary increase in the number of noisy pixels, but the increasing trend of more hot pixels is not affected by the flares.

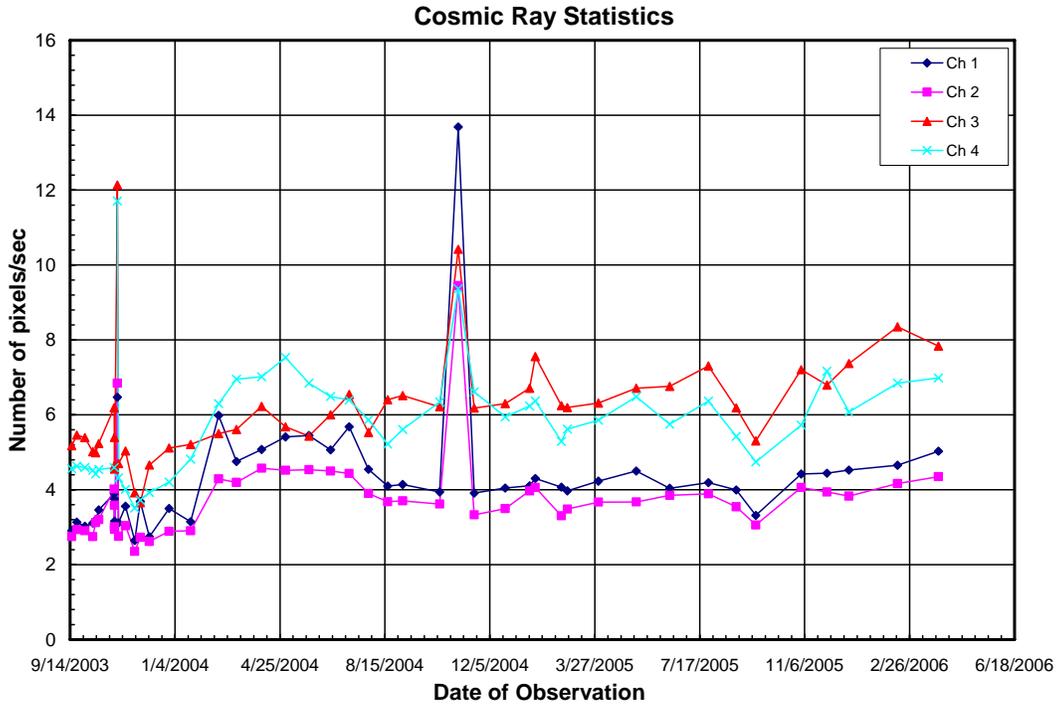

Figure 5. Cosmic ray statistics for the IRAC channels 1-4. The rates have been roughly constant for the nominal mission. The spikes at November 2003 and November 2004 are measurements that were conducted during solar flares. There is a small spike in the measurement at January 22, 2005, mainly in channel 3. That measurement occurred at the end of a large flare (see Section 3).

Table 1. IRAC Median Pixel Statistics (as of April 2006)

| IRAC Channel | Hot Pixels | Noisy Pixels | Dead Pixels | CR Rate (pixels/sec) |
|---|---|---|---|---|
| 1 | 48 | 19 | 13 | 4.0 |
| 2 | 19 | 9 | 15 | 3.7 |
| 3 | 0 | 30 | 0 | 6.2 |
| 4 | 75[†] | 44 | 1 | 5.8 |

## 2.2. Channel 4 Hot pixels

The number of hot pixels in channel 4 has apparently been rising at a steady rate of approximately 0.22 pixel/day since launch (see Figure 4), presumably due to radiation experienced in the space environment. It is not known what is causing this increase, or why channel 3 with the same detector material does not show the same effect. Channel 4 is operated at a higher applied bias than channel 3, which might partly explain the difference between the devices. At the measured rate of increase, the number of hot pixels in channel 4 will be around 500 at the expected end of the *Spitzer* cryogen lifetime. In order to assess the impact of this on the science for the remainder of the mission, the characteristics of the hot pixels were examined in the nomops campaigns 1-19[‡].

---

[†] The median for the number of channel 4 hot pixels is not meaningful since the value is rising linearly with time during the mission; see Section 2.2 and Figure 4.

[‡] See the Spitzer web pages (http://ssc.spitzer.caltech.edu) for the schedule of instrument campaigns and observing logs.

### 2.2.1. Overall changes in dark current in Channel 4

The majority of pixels have fairly stable dark current values over the nomops period. Figure 7 shows histograms of the dark current values for channel 4 at the start of nomops (November 2003) and at the end of nomops campaign 19. In the units plotted (analog/digital units, or ADU), the hot pixel cutoff is at 123. The largest difference seems to be a "shoulder" on the distribution from about 30-50 ADU, as well as a higher "tail" and more pixels at higher counts in the nomops 19 histogram.

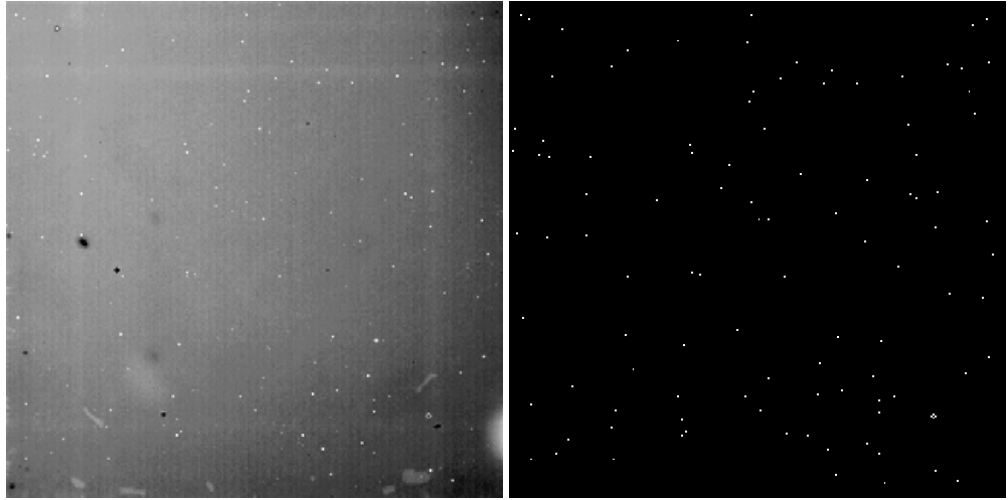

Figure 6. The channel 4 background image from campaign NOM19 is on the left. On the right is the map of hot pixels, with white indicating hot and black not.

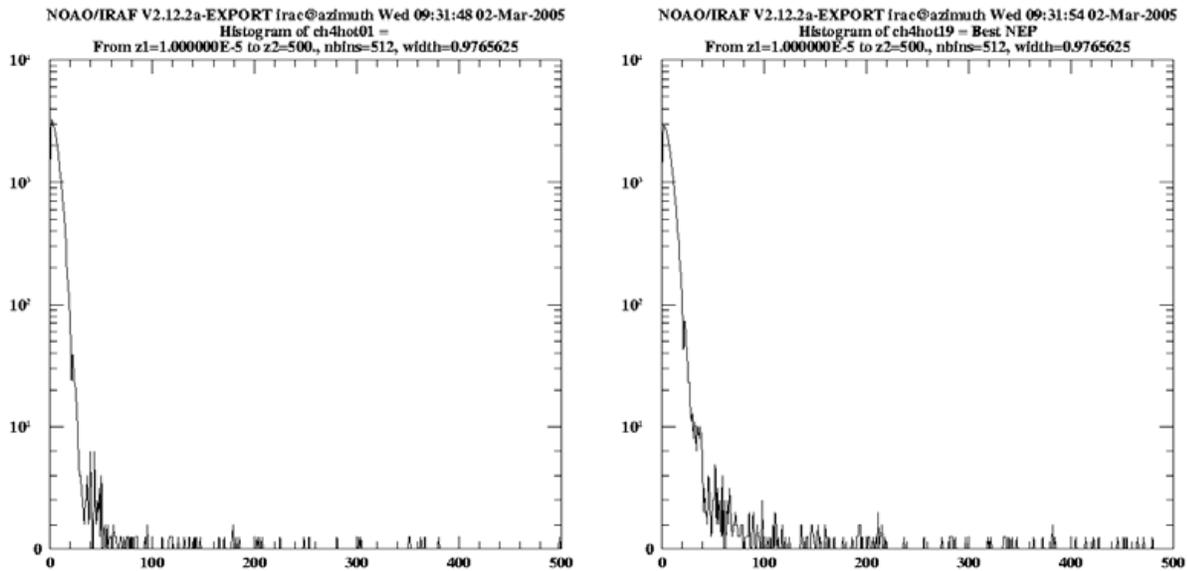

Figure 7. Histograms of dark current values for the first nomops campaign (left) and campaign 19 (right). The horizontal axis is in ADU, and the vertical axis is number of pixels.

### 2.2.2. Hot pixel generation and behavior

Increases in dark current in an individual pixel seems to happen as a step function, although some pixels seem to increase over two or three campaigns. Figure 8 shows the history of four individual pixels that transitioned from normal to hot during nomops. These pixels appeared normal until some point where the dark current jumped to a high value, where they remain more or less stable. For pixels that reach a high enough dark current value, for example pixel (90,80) in the upper left of Figure 8, the standard deviation also shows an increase at the same time.

There are also pixels that show a constant hot level throughout nomops. Presumably these were hot pixels on the ground or became hot during IOC and have remained constant (Figure 9). There are also pixels that appear to show some evolution, either taking several steps to reach a certain plateau level, or changing between different hot levels (Figure 10).

### 2.2.3. Noise characteristics of Channel 4 hot pixels

Most of the new hot pixels have levels of a few hundred ADU (Figure 7). Therefore, the noise from the increased dark current does not significantly add to the total noise in most hot pixels, since the noise is dominated by the zodiacal background in this channel, even in this low background region. Figure 11 shows the standard deviations of all pixels and the hot pixels, indicating that most hot pixels fall within the normal range of standard deviations on the array.

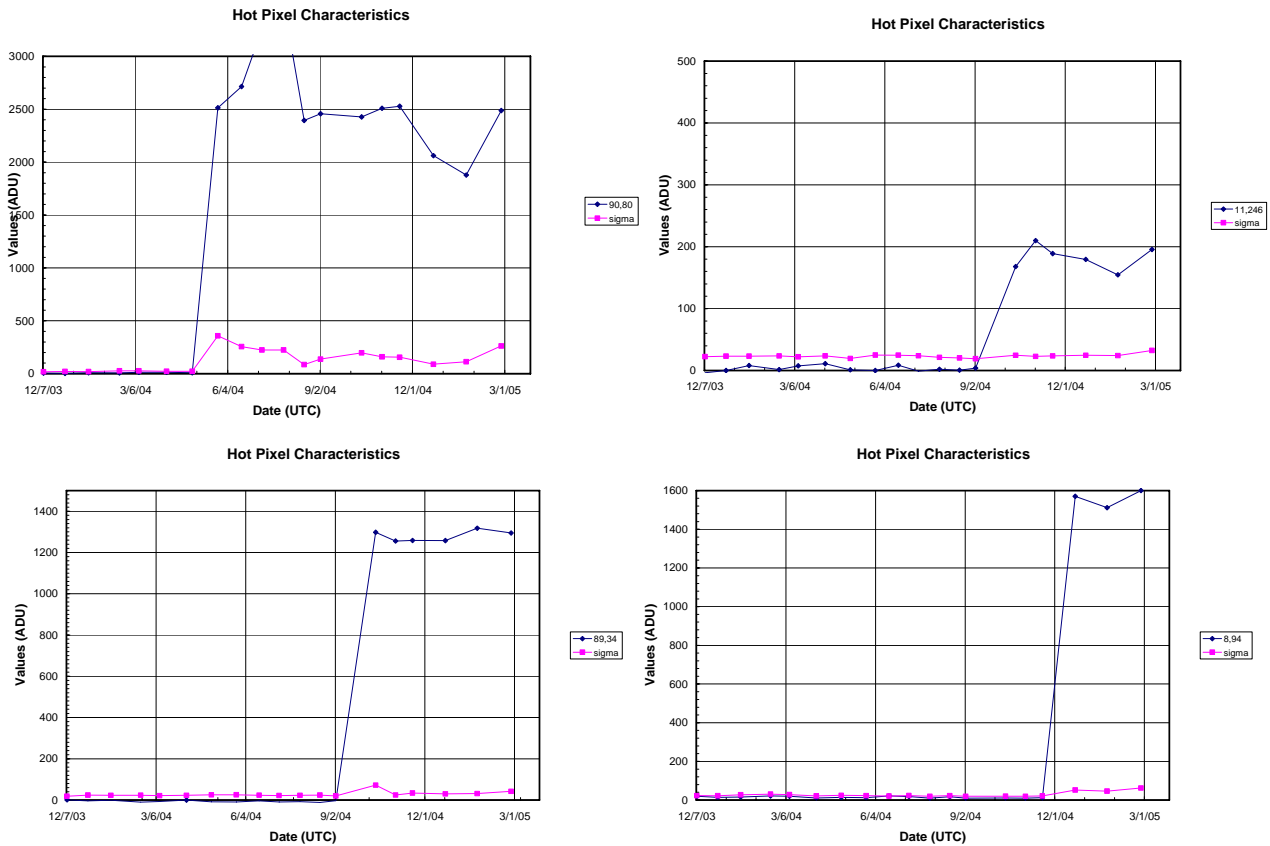

Figure 8. Plots of pixel dark current over the nominal ops campaigns for individual pixels that transitioned to "hot" status. Also shown is the standard deviation of the pixel in the dark frames.

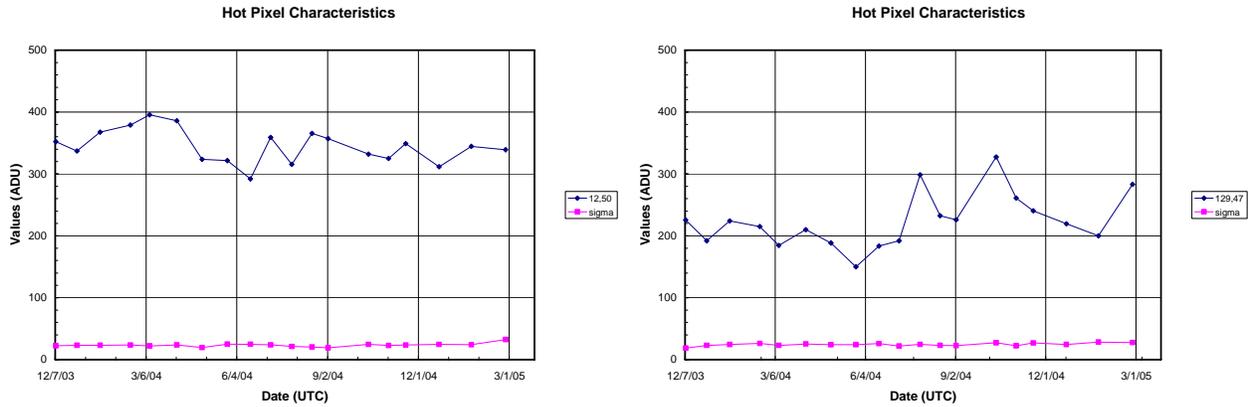

Figure 9. Hot pixels that have been hot since the start of nominal ops and have remained roughly constant.

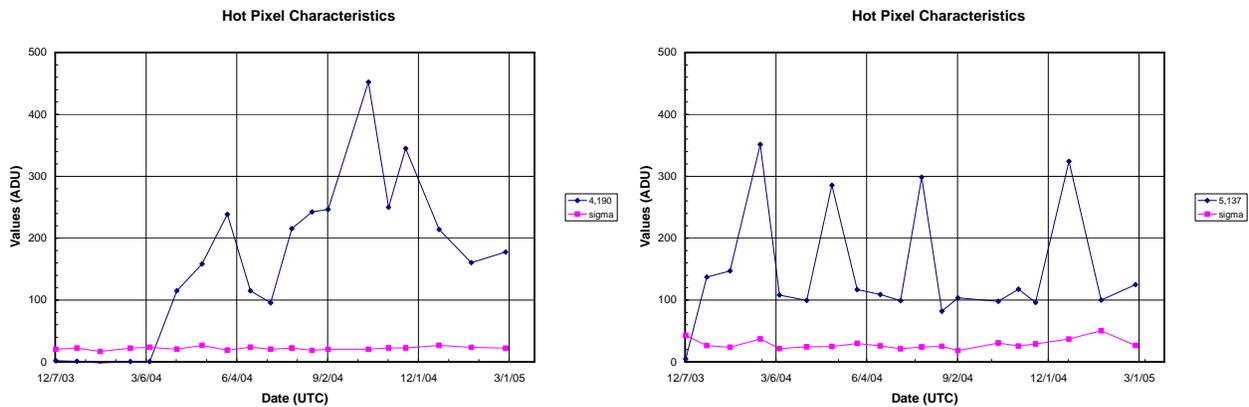

Figure 10. Pixels that have shown some evidence of non-step evolution during nominal ops. Some of the variations might be from noise in the method of obtaining the pixel statistics.

### 2.2.4. Channel 4 hot pixel discussion

The number of hot pixels in channel 4 has continued to increase since launch. The number was 160 for nomops campaign 30 (April 2006), and the current rate will put the number at 500 at the end of the cryogen lifetime. There have been occasional spikes due to solar events, but those have not had any effect on the baseline increase. Also, it seems that our current annealing strategy[1] (where the detector is heated briefly above its nominal operating temperature) is not significantly reducing the dark current of these long-term hot pixels, once a pixel is activated. The anneals do seem to remove "temporary" hot pixels caused by solar protons and other sources. The calibration and anneal strategy being used now does not allow us to easily estimate the number of hot pixels that are removed by the routine anneals, since only two dark frame reference measurements are made during a campaign, and these are after anneals. The regular anneals also have the beneficial effect of eliminating the long-term latents in channel 4.

Since the zodiacal background is relatively high in channel 4, most of the hot pixels are only slightly higher than the baseline background due to the sky, and the noise in those pixels is not significantly increased over the other pixels in the array. Therefore the pixels are for the most part still useable. The pixels do not have as much dynamic range as other lower dark current pixels, so in cases where bright sources fall on those pixels they will become saturated earlier than other pixels. However, since in general bright sources in channel 4 are fewer than in other channels, this effect is minimal.

One possible criterion for designating a bad pixel due to high dark current and noise would be that it have a noise that was greater than twice the median noise level of the whole array, so approximately 50 ADU (see Figure 11). The number of hot pixels that fall into this category is 18. Their distribution on the array is shown in Figure 12. If these pixels are generated at the same rate over the mission, we could expect to have around 80 when the cryogens are exhausted and will be mostly scattered uniformly across the field. Note that the original detector acceptance criteria allowed for a pixel operability of 99.5%, or 328 bad pixels. Therefore it appears that the observed bad pixel numbers and the growth in the numbers throughout the mission will meet requirements and will not significantly affect the quality of the observations in channel 4.

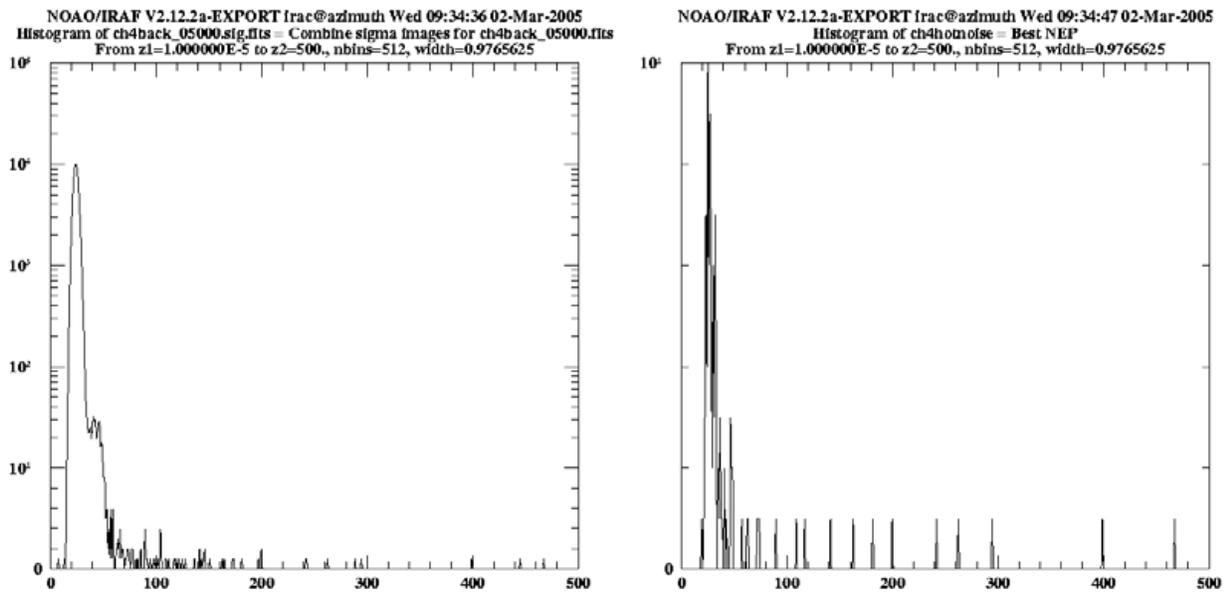

Figure 11. Histograms of the standard deviation image of the full channel 4 (left), and of only the hot pixels in channel 4 (right).

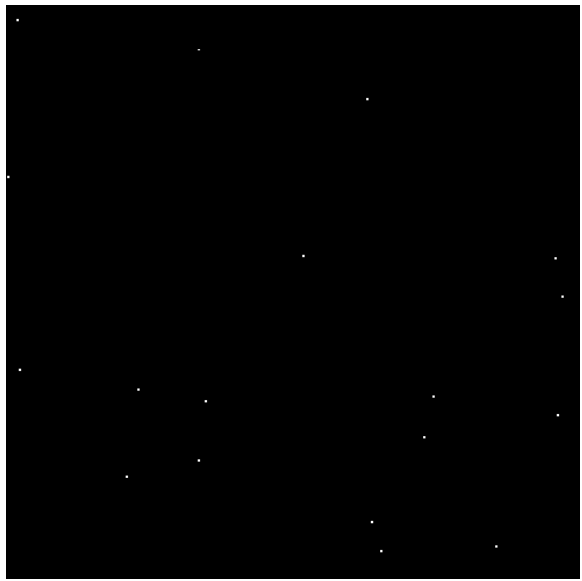

Figure 12. Distribution of hot, noisy pixels (>10 electrons/sec dark current and >2x the median standard deviation).

# 3. JANUARY 2005 SOLAR FLARE

During the IRAC Nom-18 campaign (2005 January 15 – 22; all times and dates UTC), a series of solar flares caused an increase in the proton flux measured by the GOES satellites (see Figure 13). The increased flux started after 09:00 on 2005 Jan 15, so the flare was already in progress when IRAC powered on at 18:53. The proton flux had not returned to baseline levels by the time that IRAC powered off on 2005 January 22. The >100 MeV proton flux was the highest recorded in the 29-year history of the GOES satellites, and had the fastest rise time of any event in the last 30 years[17].

The proton rates rose slowly during the first few days of the flare, with a couple local maxima, reaching a peak near 2005 January 17 15:00. The proton rates then slowly decreased over the next couple days until a sudden spike occurred on January 20, starting shortly before 07:00. The >100 MeV proton rate rose quickly to a sharp peak, and then immediately began to decay, finally returning to baseline levels around January 27.

The *Spitzer* response plan for solar flares calls for the observatory to be put into standby mode by ground command when the >100 MeV solar proton flux as measured by the GOES satellites exceeds a level of 100 proton flux units (PFUs; particles cm$^{-2}$ s$^{-1}$ sr$^{-1}$) for more than an hour. Standby mode includes powering off all science instruments. If the observatory is not downlinking data at that time, it can be several hours before the next period when commands can be uplinked. The spike on January 20 was so sudden that the process of entering standby mode could not be initiated before the flare peak had passed and the flux levels were already rapidly declining. In the end no action was taken and the IRAC observing proceeded. The data obtained during the flare are not useful for their original scientific purpose, but we can use them to assess the flare's effects on the IRAC detectors.

## 3.1. Pixel effects during the NOM-18 Campaign

Even though the flare was already in progress on power-up, the >100 MeV proton rate was not elevated, and the CR rates in the first IRAC dark frames of NOM-18 were unchanged from previous campaigns. This was expected; our experience during the 2003 October flare was that we did not see elevated CR rates until the >100 MeV proton flux was above 1 PFU, and that the observed CR rate would be CRs (pixels/sec) = 3 · (>100 MeV PFUs) plus the nominal rate.

### 3.1.1. First peak

We examined a data set from near the first flare peak (Astronomical Observation Request (AOR) 11014656 : UT 2005-01-17 13:09:52.9). We found that the CR rate was roughly 10 times nominal (43, 42, 62, 73 CR pixels/sec in channels 1 – 4, respectively). Figure 14 shows maps of pixels Jan 17 (left) and nominal images from the previous campaign (NOM-17, right), channel 1 on the top and channel 4 on the bottom. These images have the sources subtracted and scaled to show only CRs (white). The comparison is not exactly fair since the image on the left is 200 sec versus 100 sec on the right for channel 1, but the channel 4 images are the same exposure time (50 seconds). As usual, the channel 4 image CRs typically leave longer tracks, so more pixels are affected per event.

### 3.1.2. Second event

When the second event occurred, science observations were being performed, mainly 12 and 30 sec mapping AORs. The number of CR-affected pixels in a frame was determined as follows. First, a background image was constructed by taking the median of all the frames in a channel for that AOR. Then that background image was subtracted from all of the frames of the AOR. All of the pixels above a certain cutoff level were assumed to be affected. The cutoff levels were 300, 300, 400, 400 digital units in channels 1 – 4, respectively. For the BCDs which are in units of MJy/sr, the cutoff levels were 1.236, 1.556, 8.991, and 3.109 for channels 1 – 4. The number of pixels were normalized to units of pixels/sec, and plotted in Figure 15 and Figure 16.

One interesting aspect of the flare as observed by GOES-11 is that there is an initial strong peak in the proton flux, followed by a secondary peak superimposed on the exponential decline (see Figure 16). However, the CR rate, as measured by IRAC, peaks close in time to the secondary GOES peak. The IRAC rate shows only a single peak and exponential decline. As noted before, *Spitzer* is in an earth-trailing solar orbit, so it is in a different environment than the GOES satellite. Another aspect of the flare is that it occurred near a sunspot that was near the limb and rotating away from view, so *Spitzer's* exposure might have been quite different than experienced by the GOES satellites.

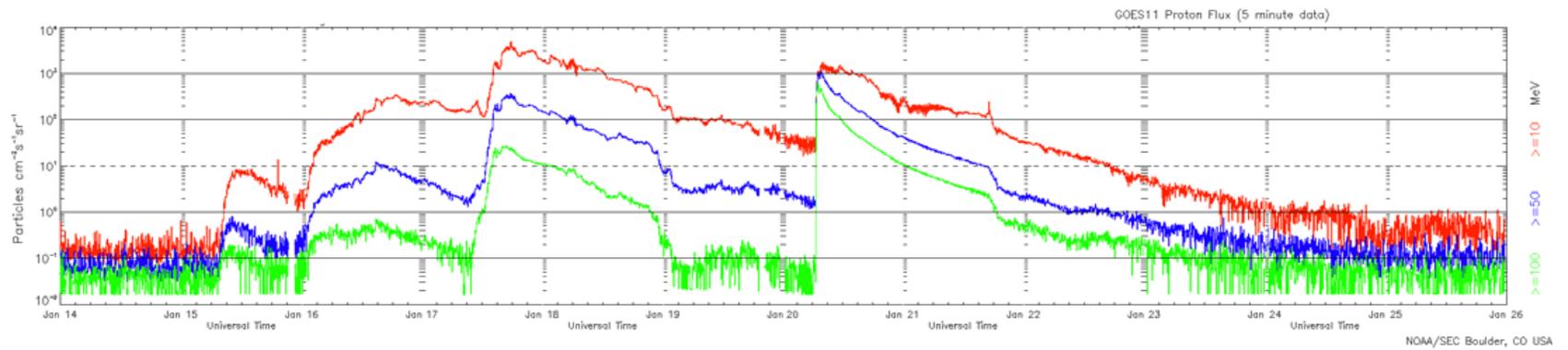

Figure 13. GOES-11 Proton flux rates from 2005 January 14 – 26. The increased proton flux started after 09:00 on 2005 January 15. The >100 MeV rates (green) did not return to baseline levels until about January 24. The lower energy protons had still not returned to baseline levels on January 26. (Figure produced by R. Arendt from plots of GOES11 data from the NOAA/SEC.)

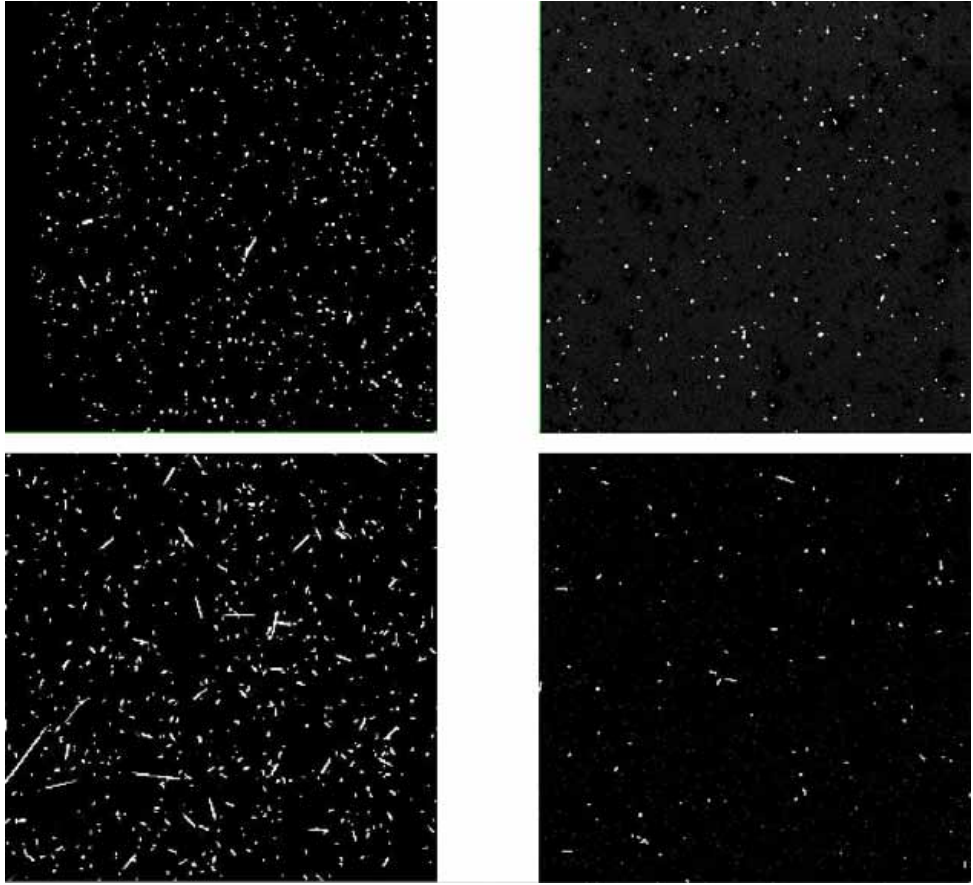

Figure 14. CR-affected pixel maps from Jan 17 (left) and nominal images from the previous campaign (Campaign NOM-17, on the right), channel 1 on the top and channel 4 on the bottom. These images have the sources subtracted and scaled to show only CRs (white). The rates are roughly 10x higher in the Jan 17 data.

There are occasional spikes in the IRAC-measured CR rates above the main trend lines; these are probably due to astronomical sources that happen to be in the field at the offset being observed. The AORs were mostly observations of the M31 galaxy, and when its large nucleus is in the field, it artificially raised the number of pixels over the CR threshold. Figure 17 shows images taken near the peak of the flare, to compare with those in Figure 14.

**3.2. Post-Flare Recovery**

**3.2.1. Anneal strategy**

For the IRAC campaign NOM-19 following the flare, it was decided that two anneals would be performed, of the type that are typically done at the start of the campaign. The purpose was to try to ensure that the arrays could be restored to their pre-flare state to prepare for science observations. Some data were taken before, between, and following the two flares to determine the effects of the anneals.

**3.2.2. Data analysis and discussion**

The data taken before and after anneals were 12 sec frames, so the usual pixel counting levels had to be slightly adjusted. The number of pixels above a cutoff value were counted as hot, and we normalized the values to the usual method from the 100 sec frames. Table 2 shows the hot pixel counts from the four channels at the start of the campaign NOM-19.

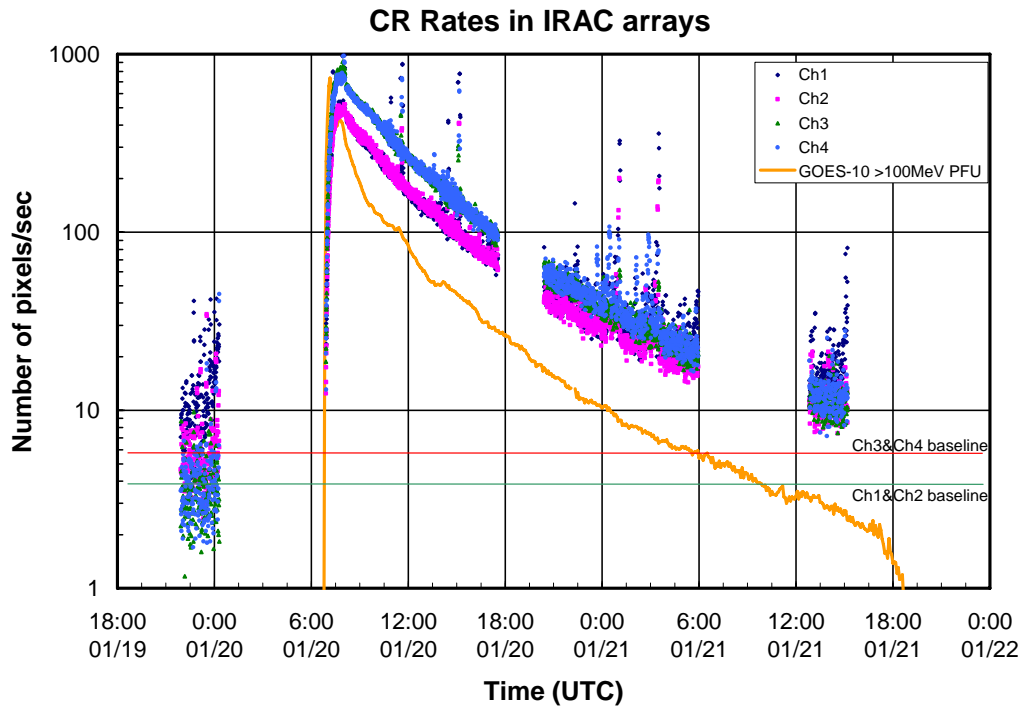

Figure 15. IRAC CR rates for the highest peak of >100 MeV protons. The occasional spikes in the IRAC CR numbers are due to astronomical sources in the field. Shown also are the baseline levels for all of the channels during nominal campaigns. The GOES-10 >100 MeV proton flux is plotted in PFUs, the CR rates are in number of pixels/sec.

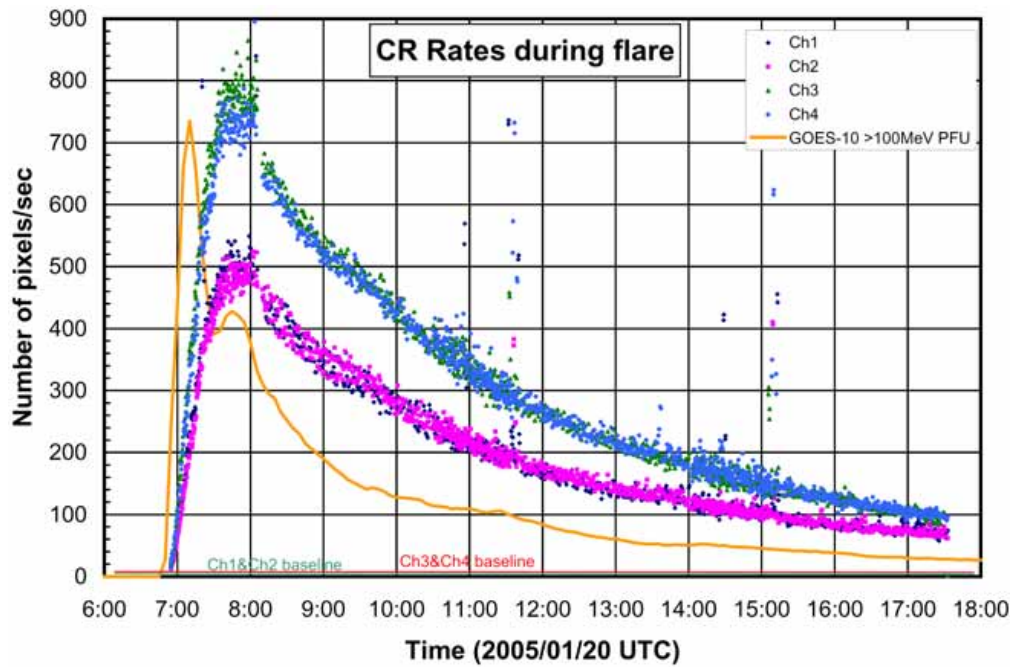

Figure 16. IRAC CR rates during the 12 hours near the highest peak of >100 MeV proton rate (this is a subset of the data shown in Figure 15). The vertical scale is now linear. The IRAC peak CR rate seems to be correlated with the secondary "bump" seen in the GOES-10 measured >100 MeV rates. The elevated CR rates did not drop off as fast as the >100 MeV proton rates.

The first anneal may have made a difference in channel 4, but apparently not in the other channels. This might not be too different from the usual situation on the first anneal, since there were several anneals after the flare during the last campaign. The hot pixel rate plot in Figure 4 shows that the hot pixel trend line was not significantly affected, and there doesn't seem to be any long-lasting effects from the flare.

It appears that the temporary effects of the flare seem to have been quickly eliminated by a repetition of the nominal power-on anneals, and the flares did not have an effect on the numbers of long-term dead or hot pixels.

Table 2. IRAC Hot Pixel Numbers After Jan 2005 Flare

| Channel | Before Anneal | Between Anneals | After Anneals |
|---|---|---|---|
| 1 | ... | 38 | 38 |
| 2 | 17 | 17 | 17 |
| 3 | 0 | 0 | 0 |
| 4 | 191 | 104 | 106 |

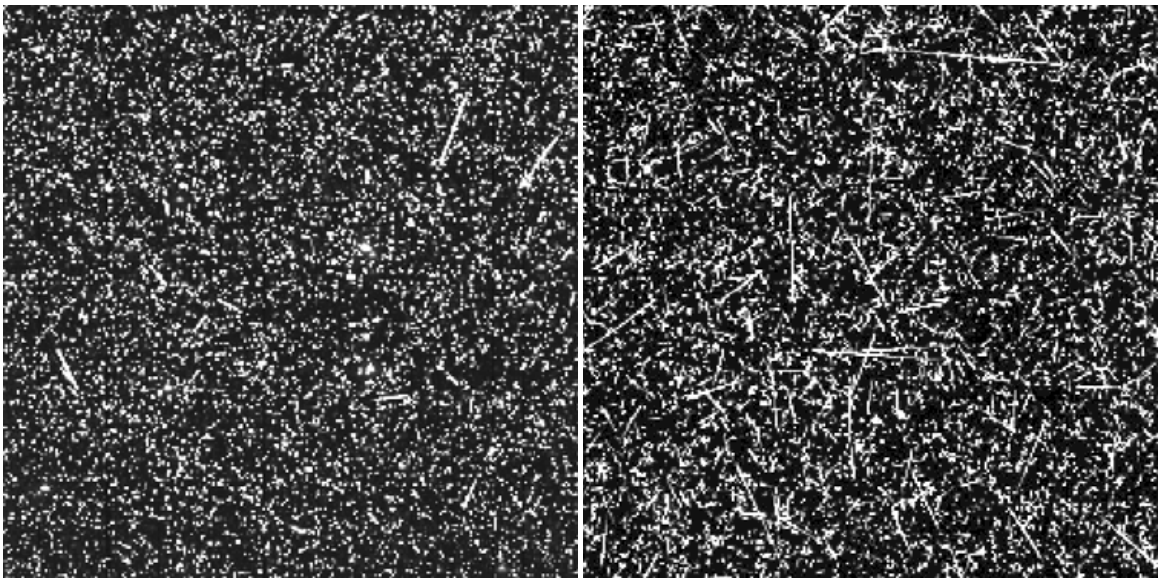

Figure 17. IRAC frames from Channel 1 (left) and Channel 4 (right) taken near the peak of the measured >100 MeV proton flux on 2005 January 20. Practically every bright (white) pixel in these images is due to the high solar proton flux.

## 4. CONCLUSIONS

The IRAC detectors have continued to perform well during the first 2.5 years of the *Spitzer Space Telescope* science mission, showing no significant changes in responsivity or array properties. The one change detected is in the number of new high dark current pixels in the channel 4 Si:As array. The number of hot pixels has been increasing steadily since launch. These new hot pixels, however, do not significantly affect the quality of the science data. The strong solar flares experienced since launch have caused temporary increases in CR rates and hot pixels, but after anneals the arrays have returned to their nominal performance with no apparent damage.

# ACKNOWLEDGEMENTS

The authors would like to thank R. Arendt for Figure 13 of this paper. Also thanks to M. L. N. Ashby for maintaining the total operational hours statistics. This work is based on observations made with the *Spitzer Space Telescope*, which is operated by the Jet Propulsion Laboratory, California Institute of Technology under NASA contract 1407. Support for the IRAC instrument was provided by NASA under contract number 960541 issued by JPL.